# Overlapped Chunked Network Coding

Anoosheh Heidarzadeh and Amir H. Banihashemi

Department of Systems and Computer Engineering, Carleton University, Ottawa, ON, Canada

July 7, 2009

*Abstract*—Network coding is known to improve the throughput and the resilience to losses in most network scenarios. In a practical network scenario, however, the accurate modeling of the traffic is often too complex and/or infeasible. The goal is thus to design codes that perform close to the capacity of any network (with arbitrary traffic) efficiently. In this context, random linear network codes are known to be capacity-achieving while requiring a decoding complexity quadratic in the message length. Chunked Codes (CC) were proposed by Maymounkov *et al.* to improve the computational efficiency of random codes by partitioning the message into a number of non-overlapping chunks. CC can also be capacity-achieving but have a lower encoding/decoding complexity at the expense of slower convergence to the capacity. In this paper, we propose and analyze a generalized version of CC called Overlapped Chunked Codes (OCC) in which chunks are allowed to overlap.[1] Our theoretical analysis and simulation results show that compared to CC, OCC can achieve the capacity with a faster speed while maintaining almost the same advantage in computational efficiency.

## I. INTRODUCTION

In the context of information flow over the packet networks, network coding is known to generally improve the network's throughput and resilience to losses [1]. A packet network can be represented as a graph, either directed or undirected, each of whose edges corresponds to a link between two network nodes. Each link is modeled as a packet erasure channel. The flow of the transmissions is called the *traffic*. Mostly in literature, the traffic is modeled by either a stochastic process or a deterministic description of the transmission instances. A problem of interest is to design codes that achieve the capacity of a network with a given traffic model.

In practice, however, the accurate modeling of the traffic is often too complex and/or infeasible [4]; namely, either the set of active nodes varies over time or the link capacities dynamically change due to the congestion or any other cross traffic phenomena. The packets may experience unconstrained delays and even be reordered. The network nodes might not be aware of successful transmissions and the feedback solution might not be appealing; because it may either contribute to additional delay or be too difficult to implement. This implies the need for *rateless* codes in the following sense: each node constructs and transmits new coded packets regardless of which packets have been received by the destination nodes. Hence the code design problem from an information-theoretic perspective translates into devising codes that perform close to the capacity for *any arbitrary* traffic.

Random linear network codes (a.k.a. dense codes) are known to be capacity-achieving over the networks with any arbitrary traffic. But this comes at the cost of computational inefficiency; each node requires $O(k)$ operations per input symbol to encode or recode $k$ symbol-long messages. The decoder also requires $O(k^2)$ operations per input symbol to invert a $k \times k$ matrix as well as $O(k)$ operations per input symbol to apply the inverse matrix to the received coded symbols. Therefore, a question from a computational complexity point of view is how to design capacity-achieving codes that are more efficient. By applying random codes on a *chunk* (a smaller sub-message of the original message), the encoding and decoding require a number of operations proportional to the size of the chunk which is smaller than $k$, the size of the original message. This is the main idea behind the design of Chunked Codes (CC) in which the message is split into chunks [4]. Each node decides which chunk(s) to transmit at any time instant using a random linear code. The code design problem has thus to deal with the following issues: how to chunk the message, and how to schedule the chunks to be coded and transmitted.

Maymounkov *et al.* [2] present a solution for designing CC that can provably achieve the capacity for any arbitrary traffic. They partition the message into non-overlapping chunks of equal size. Each node applies a random linear code to a *randomly* chosen chunk at any point in time. Each chunk has a unique content and so has to be decoded separately. By *random scheduling*, however, it may take too long for some chunks to be decoded. This reduces the speed of convergence to the capacity.

In this work, we design CC in which chunks are allowed to overlap. The decoding of one chunk, in this case, would provide some information that can help in decoding the other chunks. Consequently, the speed of convergence to capacity is increased while preserving almost the same advantage in terms of the computational efficiency. In addition to asymptotic results, we also demonstrate through finite-length simulations that when CC cannot be designed to be capacity-achieving, the application of the proposed coding scheme can reduce the overhead by about $60\%$ and $29\%$ for the line networks of length 1 and 2, respectively.

## II. PROBLEM SETUP

In this paper, we focus on the unicast network problem over line networks.[2] A line network of length $l$ is a collection of $l$ links connected in tandem whose node set includes one source, one terminal and some internal nodes. Each link is assumed to be an erasure channel. The transmitted packets may be erased with arbitrary probability and be faced with arbitrary delays. The set of successful transmissions is called a *schedule* to which the network nodes are assumed to be blind. The schedule describes paths over time on which packets can be transmitted through the network.

---

[1] Very recently, we found out that the idea of CC with overlapping chunks has also been proposed in [3] independently. Unlike this paper, however, no theoretical analysis is presented in [3].

[2] More general network scenarios can be analyzed in worst-case through the union bound analysis using the results presented here.

The schedule can be modeled by a graph specified as follows. For each network node $v$, the graph includes a set of nodes $(v, t')$ so that at each time $t'$, the node $v$ either successfully transmits or receives a packet. An edge between any two nodes $(v, t')$ and $(w, t'')$ is a *traffic edge* indicating a successful transmission initiated at node $v$ at time $t'$ and received by node $w$ at time $t''$. Without loss of generality, each traffic edge is assumed to have a unit capacity. Furthermore, an edge between any two nodes $(v, t')$ and $(v, t'')$ represents a *memory edge* with an infinite capacity. Let $s$ and $t$ denote the source and terminal nodes, respectively. The capacity of the min-cut between $(s, 0)$ and $(t, \infty)$ is called the *capacity of the schedule*. The set of paths that traverse the same set of nodes are referred to as a *flow path*. Furthermore, any line network consists of only one flow path whose number of paths equals the network's capacity. An example of a schedule with capacity 4 over a line network of length 3 is given in Figure 1.

We suppose that node $s$ is given a message of $k$ information symbols, where the symbols are strings of bits. Consider a line network with an arbitrary schedule of capacity $n$. Our problem is to design a coding scheme that achieves the capacity of the schedule, i.e., the number of information symbols, $k$, that are successfully transmitted from $s$ to $t$, asymptotically approaches $n$, as $k$ goes to infinity. We are also interested in schemes with low computational complexity.

## III. CODES ANALYSIS

We start with an overview of the results presented in [2], as related to our analysis of the overlapped chunked codes.

### A. Dense Codes

In a random linear coding scheme, each node transmits a *coded* symbol, i.e., a random linear combination of *all* previously received coded symbols. Associated with any coded symbol is a *payload vector*, i.e., the vector of coefficients that represents the mapping between the input symbols and the coded symbol. Each payload vector is assumed to be transmitted together with its coded symbol as a *packet*.

The set of received packets by the terminal with linearly independent payload vectors are called *innovative*. To decode $k$ input symbols, $k$ innovative packets are necessary and sufficient. By randomly combining packets, however, there are packets whose payload vectors are linearly dependent on the innovative packets' payload vectors. These are called *non-innovative*. The number of non-innovative packets can be investigated through a worst-case analysis. This provides a lower-bound on the number of received packets among which $k$ packets are innovative.

One can see intuitively that the probability of generating a non-innovative packet increases as the packets travel from $s$ to $t$. That is, each layer of the network adds extra dependency among the transmitted packets. Let $Q_i$, $i = 1, ..., l$, be a matrix of size $k \times n$ whose columns are the payload vectors of the $n$ packets received by node $v_i$ ($s \triangleq v_0$, $t \triangleq v_l$). For $Q_1$, all the columns have i.i.d. Bernoulli variables, i.e., the number of columns with mutual dependency is zero. As we proceed down the layers of the network, however, $Q_i$ would have some columns that are not mutually independent. We denote the collection of these columns as a sub-matrix $\mathcal{P}_i$ of $Q_i$. We also denote the columns of $Q_i \setminus \mathcal{P}_i$ which are not linearly independent as a sub-matrix $\mathcal{P}_i'$ of $Q_i$. In particular, at the terminal node, the number of innovative packets can be bounded by the sum of the number of the columns of $\mathcal{P}_l$ and $\mathcal{P}_l'$. To find a lower-bound on the capacity $n$ needed for the transmission of $k$ packets, one would need to find upper-bounds on the number of columns of $\mathcal{P}_l$ and $\mathcal{P}_l'$.

Let $M$ be a $k \times n$ matrix whose entries are each uniform Bernoulli random variables. The set of columns of $M$ that are mutually independent are called *dense*, and $M$ is called *dense* when all its columns are dense. Let $\mathcal{D}(M)$ denote the number of dense columns of $M$.

*Lemma 1:* [2, Lemma 3.8] Applying a Dense Code on a flow path with capacity $n$ and length $l$, $\mathcal{D}(Q)$ is larger than $n - l \log(ln/\epsilon)$ with probability of at least $1 - \epsilon$.[3]

The term $l \log(ln/\epsilon)$ is an upper-bound on the number of columns of $\mathcal{P}_l$. The number of columns of $\mathcal{P}_l'$ is also bounded above by $\log 1/\epsilon$ as the following shows.

*Lemma 2:* [2, Lemma 3.5] A $k \times n$ dense matrix $M$ fails to have full row-rank with probability of at most $\epsilon$, so long as $n$ is at last $k + \log(1/\epsilon)$.

Lemma 1 together with Lemma 2 give us an upper-bound of the total number of non-innovative packets at the terminal. This yields the following result for Dense Codes under arbitrary schedules.

*Theorem 1:* [2, Theorem 3.9] A Dense Code on a flow path with capacity $n$ and length $l$ fails to deliver $k$ information symbols with probability no more than $\epsilon$, so long as $n$ is at least $k + l \log(kl/\epsilon) + \log(1/\epsilon) + l + 1$. The encoding and decoding costs are each $O(k)$.[4]

### B. Chunked Codes

Chunked Codes operate by dividing the $k$ input symbols into $q$ non-overlapping chunks, each of size $\alpha = k/q$. The size of chunks is also known as the *aperture size*. Each node randomly chooses a chunk and transmits a random linear combination of received packets pertaining to the chosen chunk. Any transmission associated to the chunk $\omega \in [q]$ is called an $\omega$-*transmission* whose associated packet is an $\omega$-*packet*.

Each node $v_i$ successfully decodes $k$ input symbols so long as $Q_i$, i.e., the matrix of the received payload vectors by $v_i$

---
[3] Hereafter, the probability of decoding failure is assumed to be inversely polynomial in $k$, i.e., $\epsilon = 1/k^c$ for some positive $c$.

[4] The encoding and decoding costs are the number of operations per input symbol required for encoding and decoding tasks, respectively.

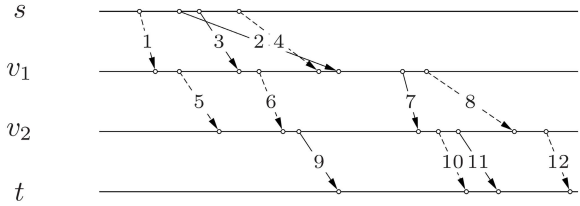

Fig. 1. A line network of length 3 with a schedule of capacity 4.

has rank $k$. Let $Q_\omega$ be $Q_i$ restricted to the columns associated with $\omega$-transmissions. Each chunk $\omega$ is decoded so long as $Q_\omega$ has rank $k/q$. This means the completion of decoding of each chunk $\omega$ is independent from that of the other chunks. By Lemmas 1 and 2, this depends on $\mathcal{D}(Q_\omega)$ and the number of $\omega$-packets at node $v_i$. We thus need to determine the number of received $\omega$-packets by each node. This quantity is a random variable because of the random nature of the scheduling and hence may deviate from its expectation. This deviation has therefore to be studied in a worst-case scenario. We give a simple example to explain this.

Suppose a CC with two chunks on the schedule depicted in Figure 1. Let us name the chunks $\mathcal{A}$ and $\mathcal{B}$. Each transmission is randomly chosen to be either a $\mathcal{A}$-transmission or a $\mathcal{B}$-transmission. The $\mathcal{A}$-transmissions and $\mathcal{B}$-transmissions are represented by solid and broken line arrows, respectively. By the definition, the min-cut capacity of the schedule restricted to the $\mathcal{A}$-transmissions (or $\mathcal{B}$-transmissions) equals the number of disjoint sets of $\mathcal{A}$-packets (or $\mathcal{B}$-packets) that form valid paths of flow from $s$ to $t$. For example in Figure 1, the $\mathcal{B}$-transmissions form a schedule of capacity 2. Furthermore, the capacity of flow of $\mathcal{A}$-transmissions is 1. The expected number of paths by $\mathcal{B}$-transmissions (or $\mathcal{A}$-transmissions) is 2. (The number of disjoint paths is 4 and this is to be divided between the two chunks.) Thus, one path is lost and this is because of those transmissions that can not be matched with any (on the upper links) to form distinct paths of flow from $s$ to $t$. These are called the *lost transmissions*.

The above example raises the key points in the analysis of CC as follows. First, the number of lost $\omega$-transmissions has to be taken into account to provide an estimate of the capacity of the flow by $\omega$-transmissions. Then, Lemmas 1 and 2 are used to derive an upper-bound of the number of non-innovative $\omega$-packets at the terminal. Finally, the total number of non-innovative packets received by node $t$ is computed by a union bound and this completes the analysis.

*Lemma 3:* For a CC over a flow path with capacity $n$ and length $l$, for each $\omega$, the $\omega$-transmissions form a flow of capacity $f$ of at least

$$\left(1 - O\left(\left((l^4 q^2/n)\log(ln/\epsilon)\right)^{1/4}\right)\right) \cdot (n/q)$$

with probability no less than $1 - \epsilon$, so long as

$$l^4 q^2 \log \frac{ln}{\epsilon} = o(n). \tag{1}$$

Lemma 3 is based on bounding the number of lost transmissions.[5] From the above, it can be deduced that $f$ asymptotically approaches its expected value, i.e., $n/q$, as long as condition (1) holds. The following is then easy to prove.

*Corollary 1:* For a CC over a flow path with capacity $n$ and length $l$, $\mathcal{D}(Q_\omega)$ (for each $\omega$) is smaller than $n/q - l(\log(n/q) + \log(1/\epsilon) + \log l + 1)$, with probability at most $\epsilon$.

Corollary 1 along with Theorem 1 result in the following for CC on arbitrary schedules.

*Theorem 2:* A CC can deliver $k$ information symbols to the terminal over a flow path with capacity $n$ and length $l$ with probability of failure no larger than $\epsilon$, so long as $n$ is at least $k + ql\log(kl/\epsilon) + q\log(1/\epsilon) + q\log q + q$, provided that $l^4 q^2 \log(kl/\epsilon) = o(k)$. Furthermore, the encoding and decoding costs are each $O(k/q)$.

## IV. FURTHER ANALYSIS

Through the results of Section III, it can be seen that the performance of CC is subject to the following factors: (1) the number of non-innovative packets, (2) the number of lost transmissions and (3) the condition of decoding completion.

The first results from the *random choices* of linear combinations. It is not always the best to *randomly* choose a subset of packets to be combined. There might be a solution to selectively combine the packets so that the coded packet is innovative with the highest probability. This is the case when each node has some knowledge of the received packets by the destination nodes. Thus, in our setup, there is no way to reduce the number of non-innovative packets.

The second is a result of the fact that each node schedules chunks *randomly* regardless of whether they form valid paths of flow. More valid paths might be packed in a given schedule by keeping track of the transmissions. But, as long as the schedule is assumed to be arbitrary, the worst-case analysis in Section III-B is an appropriate approach. Therefore, the upper-bound on the number of lost paths cannot be improved.

Last, but most important, is the decoding completion condition. Each chunk has to be decoded separately when the chunks do not overlap in any symbol, i.e., $Q$ has rank $k$ only if each $Q_\omega$ has rank $k/q$. This makes the speed of convergence to the capacity very slow. A question is whether one can devise a chunking method by which $Q_\omega$ does not necessarily have to have rank $k/q$ for $Q$ to have rank $k$. We answer this question in positive by allowing chunks to overlap. This relaxes the decoding completion condition resulting in a faster speed of convergence. The following gives the intuition behind the overlapping chunks.

Suppose the network is reduced to an erasure channel with an unknown rate of packet loss. The following results are straightforward, though they are not special cases of the results in Section III. (There is no internal node across the network and so we do not need to worry about the density loss caused

---

[5]We bring this to the attention of the reader that Lemma 3 is different from [2, Theorem 4.1], though the proofs have generally a similar structure.

by the sub-matrix $\mathcal{P}_l$ of $Q_l$.[6])

*Corollary 2:* A Dense Code fails to deliver $k$ information symbols over an erasure channel with probability of at most $\epsilon$, so long as $n > k + \log 1/\epsilon$. The coding costs are each $O(k)$.

*Corollary 3:* A CC can decode $k$ information symbols from $n$ packets received by the terminal with probability of at least $1 - \epsilon$, so long as $n$ is larger than $k + q \log 1/\epsilon + q \log q$. The encoding and decoding costs are each $O(k/q)$.

By Corollary 3, the more the chunks, the better the computational efficiency. The Chunked Codes, however, are capacity-achieving so long as the number of chunks is taken to be $o\left(k/\log k\right)$. This provides a lower-bound on the computational cost and a simple calculation shows that the encoding and decoding costs can not be reduced down to $O(\log k)$.

Recently, Studholme and Blake [6] introduced a class of erasure codes, in essence, similar to CC that operate on $k$ chunks, each of size $\alpha$ such that any two consecutive chunks overlap in all but one symbol. The chunks are scheduled in the same way as CC. It has been shown in [6] that for sufficiently large $\alpha > \sqrt{k}$, this coding scheme achieves the capacity of any erasure channel as fast as Dense Codes, i.e., with the overhead $O(\log k)$. This comes from the fact that the matrix $Q$ in this case behaves indistinguishably from the $Q$ in the dense coding schemes. This is while the $Q$ in the scheme of [6] has a structure similar to the $Q$ for CC. The encoding and decoding costs are each $O(\sqrt{k})$ and a simple comparison indicates that Chunked Codes having the same computational complexity need larger overhead of $O(\sqrt{k} \log k)$. This shows that the rank of $Q_l$ is much less dependent on the rank of each $Q_\omega$ when the chunks overlap as compared to the case with non-overlapping chunks. In the next section, we use this idea to design Overlapped Chunked Codes (OCC) which provide a better tradeoff between the speed of convergence and the coding costs.

## V. OVERLAPPED CHUNKED CODES

### A. Preliminaries

*Conjecture 1:* [6, Conjecture 4.2] Consider a $k \times n$ binary matrix $M$ whose columns are each chosen independently at random with nonzero entries restricted to lie within an aperture of sufficiently large size $\alpha > \sqrt{k}$. Then $M$ has rank less than $k$ with probability of at most $2^{-(n-k)}$.

*Corollary 4:* Let $M$ be a matrix as described above except that only $q$ apertures are to be chosen. The set of $q$ apertures are a subset of apertures in which any two contiguous apertures overlap in $\gamma$ symbols. For any $\gamma > \sqrt{k}$, $M$ fails to have full row rank with probability of at most $2^{-(n-k)}$.

A Chernoff bound shows the matrix $M$ includes $n/q$ columns pertaining to any aperture with probability of at least $1 - \epsilon$, so long as

$$q^2 \log \frac{n}{\epsilon} = o(n). \qquad (2)$$

We notice that condition (2) is a weaker version of condition (1). Corollary 4 thus brings about the following.

[6]This seems to have been overlooked in [2] when comparing CC to LT Codes [5] over an erasure channel.

*Corollary 5:* Let $M$ be a $k \times n$ binary matrix. Take $q$ apertures of size $\alpha$ such that any two contiguous apertures overlap in sufficiently large $\gamma > \sqrt{k}$ symbols. Furthermore, suppose that each aperture appears in $n/q$ columns. Then, $M$ does not have rank $k$ with probability of at most $2^{-(n-k)+\log \epsilon}$, so long as condition (2) is met.

### B. Analysis

Suppose that the input message of size $k$ is partitioned into $q$ chunks, each of size $\alpha$, and each node randomly chooses a chunk to operate on by a random linear code. Unlike CC of [2], in the proposed chunking scheme any two contiguous chunks overlap by $\gamma = \alpha - k/q$ symbols in an end-around fashion. Furthermore, in order to ensure that all the input symbols are covered equally by the chunks, $(\alpha - \gamma)$ has to be an integer divisor of $\alpha$. For the simplicity of exposition, we consider

$$\gamma = \left(\frac{\tau - 1}{\tau}\right)\alpha$$

for any divisor $\tau$ of $\alpha$ ($1 \leq \tau \leq \alpha$). Note that $\tau = 1$ implies that there is no overlap between chunks and for any $\tau > 1$, the overlap is increased as $\tau$ increases so that for $\tau = \alpha$, any two contiguous chunks overlap in all the symbols but one.

Now suppose that an OCC with the aperture size $\alpha$ and overlap $\gamma$ is applied to a flow path of capacity $n$ and length $l$. By Corollary 1, the number of dense columns associated with each chunk $\omega$ in $Q_l$ is at least $f - l \log(fl/\epsilon)$, with probability no less than $1 - \epsilon$, where $f$ denotes the capacity of the flow of $\omega$-transmissions. The concentration theorem then implies that $f$ is about $n/q$ and this fails with arbitrarily small probability if the number of chunks has been chosen with care, i.e., if condition (1) holds. Thus for any $\tau > 1$, as long as $\gamma \geq \sqrt{k}$, Corollary 5 applies.

*Theorem 3:* An OCC with $q$ chunks of size $\alpha$ and overlap $\gamma \geq \sqrt{k}$ fails to deliver $k$ information symbols on a flow path with capacity $n$ and length $l$ with probability no more than $\epsilon$, so long as

$$n > k + ql \log kl/\epsilon + ql + \log 1/\epsilon + 1,$$

provided that

$$l^4 q^2 \log(kl/\epsilon) = o(k) \qquad (3)$$

and

$$\tau = o\left(\alpha/(l^4 q \log(kl/\epsilon))\right). \qquad (4)$$

Moreover, the encoding and decoding costs are each $O(\alpha)$.

Let $\tau^*$ be the smallest divisor of $\alpha$ such that $\gamma \geq \sqrt{k}$. For now, the aperture size is assumed to be chosen so that such $\tau^*$ does exist. For any $\tau > \tau^*$, $\gamma > \sqrt{k}$, and so the above result holds. For a given aperture size, however, the larger the $\tau$, the more are the chunks, and by Theorem 3, the slower is the speed of convergence. We resolve this by setting $\tau = \tau^*$.

Now suppose there is no such $\tau^*$ for a given $\alpha$, but condition (1) still holds for some $\tau$. It can be easily seen that $Q_l$ has the structure as specified in Corollary 4 except that $\gamma < \sqrt{k}$. To the best of our knowledge, the rank property of any such matrix with $\tau \neq 1$ is an open problem. But at the very least, the rank property of the class of such matrices with $\tau = 1$ can be thought of as a lower-bound for any such matrix with $\tau \neq 1$.

*Theorem 4:* An OCC with $q$ chunks of size $\alpha$ and overlap $\gamma < \sqrt{k}$ can successfully transmit $k$ information symbols over a line network of length $l$ with an arbitrary schedule of capacity of at least $k + ql \log kl/\epsilon + \kappa$, where $\kappa \leq q \log 1/\epsilon + q \log q + q$, given that conditions (3) and (4) are met.[7] Moreover, the coding costs are each $O(\alpha)$.

### C. Concluding Remarks

Here we compare CC to OCC, both for arbitrary schedules over line networks.

*Remark 1:* The speed of convergence in both coding schemes is dominated by the largest term which is the same. For finite-length codes, however, even the smallest term does make a difference as shown in the simulations.

*Remark 2:* For a fixed number of chunks, the overlapped chunked coding scheme is capacity-achieving with larger speed of convergence. This however comes at the cost of slightly increasing the computational complexity. (The aperture size needs to become larger to provide overlap between the chunks.) That is, OCC can provide a better tradeoff between the speed of convergence and the coding costs. For a given aperture size, however, OCC asymptotically achieve the capacity with lower speed when compared to CC.

*Remark 3:* Regarding the complexity considerations in practical scenarios, suppose that the aperture size is bounded above so that none of the two coding schemes is capacity-achieving with an asymptotically vanishing overhead. OCC can be provably shown to achieve the capacity faster than CC (but the overhead per symbol does not vanish with $k$) with the same encoding/decoding complexity.

## VI. SIMULATION RESULTS

In this section, the performance of the coding schemes over the line networks are studied through simulations. We consider networks with length $l \in \{1, 2, 4\}$. The networks are each simulated with randomly generated schedules of capacity $n$, for integers $n \in \{1024, ..., 1984\}$. The number of information symbols, $k$, is set at $1024$ and each symbol is assumed to be a binary random variable. We are interested in the probability of successful decoding, i.e., the event of recovering the $k$ information symbols at the terminal, as a function of the network capacity. We would also like to investigate how this function changes with $l$, the number of chunks $q$, and the overlap $\gamma$. For different values of $n$ and $l$, we apply the coding schemes to random schedules until we have $100$ successful decoding events.

---
[7]The upper bound on $\kappa$ is tight for $\tau = 1$, though it becomes looser as the value of $\tau$ increases.

In the first experiment, we set $q = 2$ or $4$ for CC and $4$ for OCC. This is to stay faithful to conditions (3) and (4). Figure 2 shows the results for $l = 4$. The asymptotic upper bounds are also given for reference. In general, our results for different values of $l$ show that OCC is always superior to CC (this can also be seen in Figure 2). We notice that it should not be surprising that the finite-length results in Figure 2 surpass the asymptotic upper bounds, as the former is an average value while the latter is obtained based on a worst-case analysis.

In the second experiment, fixing the computational complexity (for a fixed aperture size, e.g., $\alpha = 16$), we study the performance of OCC with different overlaps. This will result in larger values of $q$, too large to comply with conditions (3) and (4). We set $q = 64$ for CC and $q \in \{128, 256, 512, 1024\}$ for OCC. Figure 3 shows the advantage of OCC in reducing the overhead. It can also be seen (in our simulations) that there is always an optimal $\gamma \neq 0$ which maximizes the probability of successful decoding.

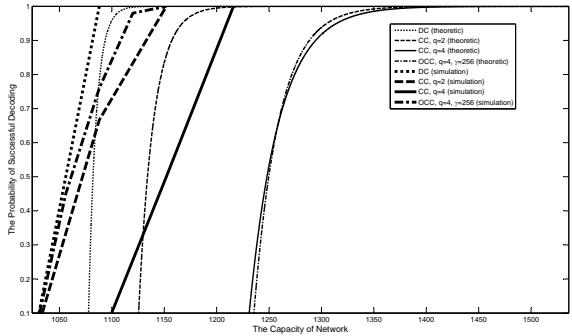

Fig. 2. The performance comparison of Dense Codes (DC), CC and OCC over line networks of length 4.

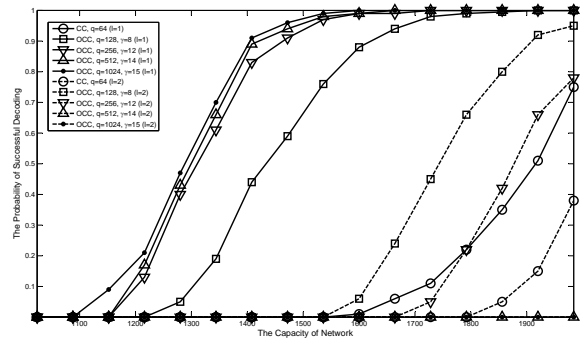

Fig. 3. The performance comparison of CC and OCC with $\alpha = 16$ over line networks of length 1 and 2.